\documentclass[aps,prd, nofootinbib]{revtex4}

\usepackage{amsmath, amsfonts, amsthm, amssymb, graphicx}
\usepackage{subfig}
\usepackage{epstopdf}

\usepackage{fontenc}
\usepackage{textcomp}

\def\be{\begin{equation}}
\def\ee{\end{equation}}
\def\ba{\begin{eqnarray}}
\def\ea{\end{eqnarray}}

\def\bdm{\begin{displaymath}}
\def\edm{\end{displaymath}}

\def\bq{\begin{quote}}
\def\eq{\end{quote}}

\def\del{\partial}
\def\ltap{\ \raise.3ex\hbox{$<$\kern-.75em\lower1ex\hbox{$\sim$}}\ }
\def\gtap{\ \raise.3ex\hbox{$>$\kern-.75em\lower1ex\hbox{$\sim$}}\ }
\def\gl{\ \raise.5ex\hbox{$>$}\kern-.8em\lower.5ex\hbox{$<$}\ }
\def\roughly#1{\raise.3ex\hbox{$#1$\kern-.75em\lower1ex\hbox{$\sim$}}}

 at 10truept

\newcommand{\beq}{\begin{equation}}
\newcommand{\eeq}{\end{equation}}
\newcommand{\bea}{\begin{eqnarray}}
\newcommand{\eea}{\end{eqnarray}}
\newcommand{\beqa}{\begin{eqnarray}}
\newcommand{\eeqa}{\end{eqnarray}}
\newcommand{\nn}{\nonumber\\}
\newcommand{\ud}{\mathrm{d}}

\def \pd {\partial}

\begin{document}

\title{Bi-galileon theory II: phenomenology}

\author{Antonio Padilla} 
\email[]{antonio.padilla@nottingham.ac.uk}
\author{Paul M. Saffin} 
\email[]{paul.saffin@nottingham.ac.uk}
\author{Shuang-Yong Zhou} 
\email[]{ppxsyz@nottingham.ac.uk}

\affiliation{School of Physics and Astronomy, 
University of Nottingham, Nottingham NG7 2RD, UK} 

\date{\today}

\begin{abstract}
We continue to introduce bi-galileon theory, the generalisation of the single galileon model introduced by Nicolis {\it et al}. The theory contains two coupled scalar fields and is described by a Lagrangian that is invariant under Galilean shifts in those fields. This paper is the second of two, and focuses on the phenomenology  of the theory. We are particularly interesting in models that admit solutions that are asymptotically self accelerating or asymptotically self tuning. In contrast to the single galileon theories, we find  examples of self accelerating models that are  simultaneously free from ghosts, tachyons and tadpoles, able to pass solar system constraints through Vainshtein screening, and do not suffer from problems with superluminality, Cerenkov emission or strong coupling. We also find self tuning models and discuss how Weinberg's no go theorem is evaded  by breaking Poincar\'e invariance in the scalar sector. Whereas the galileon description is valid all the way down to solar system scales for the self-accelerating models, unfortunately the same cannot be said for self tuning models owing to the scalars backreacting strongly on to the geometry.  

\end{abstract}


\maketitle

\section{Introduction}
Galileon theory was  developed by Nicolis {\it et al} \cite{gal}, in order to facilitate a model independent study of certain infra-red modifications of gravity.  They considered a class of scalar tensor theories of gravity, where all modifications of General Relativity are encoded in the Lagrangian for a single scalar field propagating in Minkowski space. The scalar field Lagrangian ${\cal L(}\pi, \del \pi,  \del\del  \pi)$ is  invariant under a {\it Galilean} symmetry, $\pi \to \pi+b_\mu x^\mu+c$.  The inspiration for the galileon description comes from co-dimension one brane world models exhibiting infra-red modifications of gravity \cite{dgp, kogan, ruth, me}.  In these models, gravity on the brane is mediated by the exchange of  the graviton and an additional scalar, often corresponding to the strongly coupled brane bending mode \cite{luty}.   This strong coupling allows us to take a  non-trivial limit in which the graviton and the scalar decouple in the $4D$ effective theory \cite{nicolis}. The scalar sector contains higher order self interactions and is Galilean invariant,  a remnant of Poincar\' e invariance in the original bulk spacetime. Many features of the original brane models such as self-acceleration \cite{sa}, instabilities \cite{saghosts} and Vainshtein effects  \cite{dgp-vainsh,  dvalithm} can be studied at the level of the corresponding galileon theory \cite{gal, clare, sami, oriol}.

In our companion paper \cite{otherpaper}, we introduced {\it bi-galileon theory}. This  extends Nicolis {\it et al}'s model to two coupled galileon fields (see \cite{fairlie, pforms, solitons, wesley} for further extensions). Bi-galileon theory has particular relevance to co-dimension {\it two} brane world models exhibiting infra-red modifications of gravity \cite{casc1, casc2, casccos, cline, SLED, koyama, nk, ccpap, zammo}. Indeed, in \cite{otherpaper}, we showed that the boundary effective field theory for the cascading cosmology model \cite{casc1, casc2, casccos} corresponds to a bi-galileon theory in the decoupling limit. In an orthogonal paper \cite{solitons}, we considered a multi-galileon extension with internal symmetries, using the higher order interactions to evade Derrick's theorem and  stabilise soliton solutions. However, in this paper, we shall  return to the galileon as a means of modifying gravity.

The most general bi-galileon theory \cite{otherpaper, fairlie, pforms}  corresponds to a Lagrangian for two coupled scalar fields, $\pi$ and $\xi$, propagating on Minkowski space. The Lagrangian is bi-galilean invariant, in that it remains unchanged under the following transformation $\pi\to \pi+b_{\mu}x^{\mu}+c\,,\quad \xi\to \xi+\tilde b_{\mu}x^{\mu}+\tilde c$ . By coupling one of the scalars directly to matter, through the trace of the energy-momentum tensor, we can interpret this as a modified theory of gravity mediated by the usual graviton plus  two additional scalar fields. As in \cite{gal}, we neglect any direct mixing between the graviton and the scalars, although we do include mixing between the scalars. This is consistent with the decoupling limit of the boundary effective theory for the cascading cosmology model. Indeed, we argued that this should also apply to any co-dimension two braneworld model exhibiting infra-red modifications of gravity.

Although our previous paper \cite{otherpaper} was largely devoted to the formulation of our theory, we did initiate a study of maximally symmetric vacua, establishing elegant geometric techniques for assessing their stability. We did not, however, look at the phenomenological properties of these vacua. This is the subject of this paper. We will be particularly interested in two types of vacua: {\it self-accelerating} vacua and {\it self tuning} vacua. Roughly speaking, a {\it self accelerating} vacuum is one which undergoes cosmic acceleration even in the absence of any vacuum energy. There has been plenty of interest in these scenarios recently \cite{sa}, since they represent a gravitational alternative to dark energy \cite{de}. We should be clear that self-acceleration does not go as far as solving the cosmological constant problem. One has to assume the existence of some unknown mechanism that sets the cosmological constant to zero, and then argue that  the observed acceleration \cite{acceleration} can be explained by new gravitational physics kicking in on cosmological scales.

Self-accelerating vacua are often plagued by ghost-like instabilities \cite{saghosts}. One of the aims of the original galileon paper \cite{gal} was to see if self-acceleration is possible  without ghosts. For a single galileon, ghost-free self acceleration {\it is} possible although to avoid overly restricting the domain of validity of the theory one must include a tadpole term which essentially renormalises the vacuum energy to non-zero values \cite{gal}. In section \ref{sec:selfacc} we will study self-accelerating vacua in our bi-galileon theory, demonstrating the fact that ghost-free self acceleration is possible without  any of the additional problems found in the single galileon theory. 

A {\it self tuning} vacuum is one which is insensitive to the vacuum energy. Such a vacuum is Minkowski even in the presence of a large vacuum energy. Self tuning mechanisms should, in principle, solve the old cosmological constant problem by dynamically tuning the vacuum curvature to zero whatever the vacuum energy.  Whilst Weinberg's no-go theorem makes self tuning impossible for constant field configurations, it does not rule out more general scenarios \cite{nogo}. Indeed, co-dimension two braneworld models offer some hope for developing a successful self-tuning mechanism \cite{casc1, casc2, casccos, cline, SLED, koyama, nk, ccpap}. The reason is that adding vacuum energy to a co-dimension two brane merely alters the bulk deficit angle and not the brane geometry. Difficulties arise when one tries to study  nontrivial branes geometries as this can sometimes introduce problems with singularities \cite{cline} and perturbative ghosts~\cite{higherdgp}.  Furthermore, on a technical level, going beyond the vacua, even perturbatively, can be quite challenging in co-dimension two models. 

Given its association with co-dimension two braneworld models, it is no surprise that our general bi-galileon theory admits self tuning solutions. Indeed,  as we will see in section \ref{sec:selftun}, the corresponding vacua can sometimes be stable. For both self tuning and self accelerating solutions, the study of non-trivial excitations is not too demanding. We will be particularly interested in spherically symmetric excitations so we can compare our results to observations within the solar system. Typically one would expect the additional scalar fields to ruin any agreement with solar system tests of gravity, mainly through the troublesome vDVZ discontinuity \cite{vdvz}. For both self acceleration and self tuning, we require the scalars to have an order one effect at cosmological scales and need some mechanism to suppress this on solar system scales. To achieve this we need to appeal to  non-linear effects. Two possible mechanisms
spring to mind: the chameleon mechanism \cite{cham}, which relies on non-linear coupling to matter, and the Vainshtein mechanism \cite{vainshtein, dgp-vainsh}, which relies on non-linear self interactions. Since we have assumed that there is a linear coupling to matter we must rely on the Vainshtein mechanism and  scalar-scalar interactions. As we will see in section \ref{sec:selfacc}, for self acceleration, this will help screen the scalars at short distances, so that our theory passes solar system constraints. In contrast, for self-tuning, we will see in section \ref{sec:selftun} that one cannot ``self tune away" a large vacuum energy and still hope to pass solar system constraints, at least at the level of the galileon description. This is because the backreaction of the scalars onto the geometry  is too large and so the galileon description breaks down.

Given a  solution, our priority has been  to establish whether it is perturbatively stable, and if so, whether it admits spherically symmetric solutions that can pass solar system tests of gravity through Vainshtein screening. For self-acceleration, this was in fact possible, but there are, of course, other things to consider. For example, what is the cut-off for  our theory?  This is particularly relevant because the Vainshtein mechanism is closely linked to strong coupling effects \cite{dvalithm}. In DGP gravity, for example, fluctuations around the Minkowski vacuum become strongly coupled at length scales below around $1000$ km, corresponding to a rather low momentum cut-off in the effective $4D$ theory~\cite{luty}. However, one can argue that we do not live in the vacuum, so this is not necessarily a problem \cite{nicolis}. The relevant cut-off comes from considering fluctuations about the non-trivial spherically symmetric solution arising in the solar system. This can sometimes be much higher, as is the case for the asymptotically flat DGP solution \cite{nicolis}. Of course, we need to check this explicitly on a case by case basis. As the background solution changes with scale, so the strong coupling scale will run. There exists a critical radius at which quantum effects start to dominate and one cannot trust the classical background. We will require this scale to be shorter that the Schwarzschild radius of the Sun, below which we don't expect to be able to trust the galileon description anyway.
 
Fluctuations about spherically symmetric solutions can also suffer from both superluminal and extreme subluminal propagation of fields. The latter can cause a large amount of Cerenkov emission, spoiling the quasi-static approximation of the solution \cite{gal}. Both of these effects were found to be problematic for the single galileon theory, but can be avoided {\it simultaneously} in the bi-galileon theory. Indeed, despite placing quite a few constraints on our theory, we will find that  in some cases we tick all the boxes. This is in contrast to the case of a single galileon for which Cerenkov emission and a low scale of strong coupling can only be avoided in a ghost-free theory by introducing a tadpole. For self-acceleration, such a tadpole is theoretically undesirable as all it really does is renormalise the vacuum energy to a non-zero value. To further avoid superluminality problems, one is forced to eliminate all interactions making it impossible to satisfy solar system constraints. None of this is an issue in bi-galileon gravity -- even {\it without} a tadpole, {\it all} of the would-be pathologies can be {\it simultaneously} avoided, including superluminality whilst retaining some higher order interactions and Vainshtein screening. This is really the main result of this paper: bi-galileon gravity can give rise to ghost free self acceleration without introducing any of the other pathologies that plagued the single galileon theory. This suggests that bi-galileon inspired cosmological models are worth developing further as a viable alternative to dark energy.
 
\section{The bi-galileon model} \label{sec:themodel}
Let us begin by reviewing the main results from our companion paper \cite{otherpaper}, with some additional niceties. We considered a modified theory of gravity with two additional scalar fields.  We treat the theory as an effective theory in Minkowski space with the following field content: a  single massless graviton, $\tilde h_{\mu\nu}$, and two scalar galileons, $\pi$ and $\xi$.  Aside from the coupling to the energy momentum tensor, $T_{\mu\nu}$, we neglect any interactions involving the graviton. Since we assume a linear coupling between the scalars and matter, we can assume, without further  loss of generality, that just one of the scalars, $\pi$, say, couples to the trace of the energy-momentum tensor, $T=\eta^{\mu\nu}T_{\mu\nu}$. If the $\pi$-matter coupling is extremely weak, there is very little deviation from General Relativity, rendering the theory uninteresting on cosmological scales.  In any event, the general theory is given by
\be \label{act}
S\left[\tilde h_{\mu\nu}, \pi,\xi; \Psi_n\right]=\int d^4 x -\frac{M_{pl}^2}{4}\tilde h^{\mu\nu} {\cal E} \tilde h_{\mu\nu}+\frac{1}{2} \tilde h_{\mu\nu} T^{\mu\nu}+{\cal L}_{\pi, \xi}+\pi T
\ee
where $\Psi_n$ are the matter fields,  ${\cal E}\tilde h_{\mu\nu}=-\frac{1}{2}\Box_4 \left(\tilde h_{\mu\nu}-\frac{1}{2} \tilde h \eta_{\mu\nu} \right)+\ldots$ is the linearised Einstein tensor, and the galileon Lagrangian is 
\be
{\cal L}_{\pi, \xi}=\sum_{0 \leqslant m+n \leqslant 4} (\alpha_{m, n} \pi+\beta_{m, n} \xi) {\cal E}_{m, n}
\ee
with
\be \label{eommn}
\mathcal{E}_{m,n}= (m+n)! \delta^{\mu_1}_ {[\nu_1} \ldots \delta^{\mu_m} _{\nu_m}
             ~\delta^{\rho_1} _{\sigma_1} \ldots \delta^{\rho_n}_{ \sigma_n]}
           (\pd_{\mu_1}\pd^{\nu_1}\pi) \ldots  (\pd_{\mu_m}\pd^{\nu_m}\pi)
             ~(\pd_{\rho_1}\pd^{\sigma_1}\xi) \ldots  (\pd_{\rho_n}\pd^{\sigma_n}\xi)\,.
\ee
We see from the matter coupling that the {\it physical} metric  is given by $g_{\mu\nu}=\eta_{\mu\nu}+h_{\mu\nu}$, where  $h_{\mu\nu}=\tilde h_{\mu\nu}+2\pi \eta_{\mu\nu}$. Given a source $T_{\mu\nu}$, $\tilde h_{\mu\nu}$ gives the usual perturbative GR solution, and so $2\pi \eta_{\mu\nu}$ gives the modified gravity correction. The field equations for the scalars are
\be
T+\sum_{0 \leqslant m+n\leqslant 4} a_{m,n}\mathcal{E}_{m,n}=0,\qquad 
\sum_{0 \leqslant m+n\leqslant 4} b_{m,n}\mathcal{E}_{m,n}=0, 
\ee
where $a_{m, n}$ and $b_{m,n}$ are given by
\be
a_{m,n}=(m+1)(\alpha_{m,n}+\beta_{m+1,n-1}), \qquad
b_{m,n}=(n+1)(\beta_{m,n}+\alpha_{m-1,n+1})\,.
\ee
and satisfy
\be
\label{eq:abRelation}
na_{m-1,n}=mb_{m,n-1}.
\ee
In \cite{otherpaper}, we initiated a study of maximally symmetric vacua, where the background stress energy tensor can have a non-zero vacuum energy $\bar T^\mu_\nu=-\sigma \delta^\mu_\nu$. For de Sitter space with Hubble radius $H^{-1}$, we can expand the metric about the here ($\vec x =0$) and now  ($t =0$), so that  for $|\vec x| \ll H^{-1}$ and  $|t| \ll H^{-1}$ we have $h_{\mu\nu}=-\frac{1}{2}H^2 x_\alpha x^\alpha \eta_{\mu\nu}$. Motivated by this we took our background fields to be 
\be \label{bg}
\bar \pi(x)=-\frac{1}{4}k_\pi x_\mu x^\mu, \qquad \bar \xi(x)=-\frac{1}{4}k_\xi x_\mu x^\mu 
\ee
where $k_\pi=H^2-H^2_{GR}$ is the difference between the Hubble parameters calculated in the modified gravity theory and in GR, and $k_\xi$ is a constant. Now a particularly interesting quantity is the action polynomial,
\be \label{actpoly}
L(k_\pi, k_\xi)=4\sigma k_\pi-\sum_{0 \leqslant m+n\leqslant 4}(\alpha_{m,n}k_\pi+\beta_{m, n} k_\xi) \left(-\frac{1}{2}\right)^{m+n} \frac{4!}{(4-m-n)!}k_\pi^m k_\xi^n  \,.
\ee
which is closely related to the on-shell action for the background galileon fields,
\be
S\left[ \bar \pi,\bar \xi; \sigma \right]=\int d^4 x \left[\sum_{0 \leqslant m+n \leqslant 4} (\alpha_{m, n} \bar \pi+\beta_{m, n}\bar  \xi) \bar {\cal E}_{m, n}\right]+ \bar \pi (-4\sigma)=\frac{1}{4}\left(\int d^4 x~x_\mu x^\mu\right)L(k_\pi, k_\xi)\,,
\ee
As we showed in \cite{otherpaper}, a stable ghost-free vacuum is one that corresponds to a local minimum of the action polynomial. In other words, we require $\frac{\pd L}{\pd k_\pi}=\frac{\pd L}{\pd k_\xi}=0$,  for a solution. Furthermore, for the solution to be ghost free we require the Hessian, $Hess (L)_{ij}=\frac{\pd^2L}{\pd k_i \pd k_j}$, to have {\it non-negative} eigenvalues.

The stability conditions for the vacuum were found by considering perturbations on the background solution. We found  that by shifting the fields $\pi \to \bar \pi+\pi, ~\xi \to \bar \xi+\xi$, we get the same form for the field equations, but with different coefficients,
\be
\sum_{1 \leqslant m+n\leqslant 4} a'_{m,n}\mathcal{E}_{m,n}=-\delta T, \qquad \sum_{1 \leqslant m+n\leqslant 4} b'_{m,n}\mathcal{E}_{m,n}=0\,, \label{eomxids}
\ee
As the vacuum is already a solution, there is no contribution from $m=n=0$.  For  $1 \leqslant m+n \leqslant 4$, these new  coefficients are related to the original ones by the linear map
\be
a'_{m, n}=\sum_{i=m}^4\sum_{j=n}^4 M_{m, n}{}^{i, j}a_{i, j}\,, \qquad b'_{m, n}=\sum_{i=m}^4\sum_{j=n}^4M_{m, n}{}^{i, j}b_{i,j}\,, \label{map}
\ee
where 
\be \label{map2}
M_{m, n}{}^{i, j}=\left(-\frac12\right)^{i+j-m-n} \left( \begin{array}{c} i \\ m\end{array}\right) \left( \begin{array}{c} j \\ n\end{array}\right) \frac{(4-m-n)!}{(4-i-j)!}k_\pi^{i-m}k_\xi^{j-n}, \qquad \left( \begin{array}{c} i \\ m\end{array}\right)=i!/m!(i-m)!
\ee
For $N<0$, we extend the definition of ``factorial" using the Gamma function, $N!=\Gamma(N+1)$, and recall that $1/\Gamma(N+1)=0$ when $N$ is a negative integer. It immediately follows that we only get contributions from $i+j \leqslant 4$.

Given that the field equations have exactly the same form in the perturbed theory, it is clear that the effective action describing perturbations also has the same form,
\be \label{pertact}
S'\left[\tilde h_{\mu\nu},  \pi,  \xi,; \Psi_n\right]=\int d^4 x -\frac{M_{pl}^2}{4}\tilde h^{\mu\nu} {\cal E} \tilde h_{\mu\nu}+\frac{1}{2} \tilde h_{\mu\nu} \delta T^{\mu\nu}+\left[\sum_{1 \leqslant m+n \leqslant 4} (\alpha'_{m, n} \pi+\beta'_{m, n} \xi) {\cal E}_{m, n}\right]+\pi \delta T
\ee
where the coefficients in the equations of motion are related to the coefficients in the effective action via the standard relations
\be
a'_{m,n}=(m+1)(\alpha'_{m,n}+\beta'_{m+1,n-1}), \qquad b'_{m,n}=(n+1)(\beta'_{m,n}+\alpha'_{m-1,n+1})\,, \label{relpr}
\ee
Note that we have also made a shift in the graviton field $\tilde h_{\mu\nu} \to  \tilde h^\text{background}_{\mu\nu}+\tilde h_{\mu\nu}$. When we study spherically symmetric solution in section \ref{sec:selfacc:vain}, we emphasise that we are working with the perturbed theory given by Eq. \ref{pertact}.

Given an action polynomial for the background, we can reconstruct the corresponding action for the theory by computing the following coefficients:
\bea
a_{m,n}=(m+1)(\alpha_{m, n}+\beta_{m+1,n-1}) &=&-(-2)^{m+n}\frac{(4-m-n)!}{4!m!n!}~\frac{\pd^{m+n+1} L}{\pd k_\pi^{m+1} \pd k_{\xi}^{n}}\Bigg|_{k_\pi=k_\xi=0}+4\sigma \delta^{m}_0\delta_0^n \label{mixed3}\\
b_{m,n}=(n+1)(\beta_{m,n}+\alpha_{m-1,n+1})=  &=&-(-2)^{m+n}\frac{(4-m-n)!}{4!m!n!}~\frac{\pd^{m+n+1} L}{\pd k_\pi^{m} \pd k_{\xi}^{n+1}}\Bigg|_{k_\pi=k_\xi=0} \label{mixed4}
 \eea
where $m, n =0, 1, 2, \ldots$, and we recall that we define $\alpha_{-1, n}=\beta_{m, -1}=0$. Note that since $\pi {\cal E}_{m-1, n}-\xi {\cal E}_{m, n-1}$ is a total derivative for $n, m \geq 1$, we are free to set, say,   $\beta_{1,n}=\beta_{2,n}=\beta_{3,n}=\ldots=0$, without loss of generality. However, it is clear from Eq. \ref{mixed3} that this choice  subsequently fixes $\alpha_{m, n}$ uniquely.

The action polynomial can also be used to quickly calculate the equations of motion for fluctuations on a solution, given by Eqs. \ref{eomxids}. Given the action polynomial, we can compute the coefficients in the perturbation equations directly, using the following formulae, valid for $1 \leqslant m+n \leqslant 4$
\bea
a'_{m, n} &=&-(-2)^{m+n}\frac{(4-m-n)!}{4!m! n!}~\frac{\pd^{m+n+1} L}{\pd k_\pi^{m+1} \pd k_{\xi}^{n}} \label{apr}\\
b'_{m, n} &=&-(-2)^{m+n}\frac{(4-m-n)!}{4!m! n!}~\frac{\pd^{m+n+1} L}{\pd k_\pi^{m} \pd k_{\xi}^{n+1}} \label{bpr}
 \eea

We conclude our review with a word or two about backreaction of the scalars on to the geometry. In \cite{otherpaper}, we argued that this backreaction could be neglected provided 
$ T_\text{scalar}^{\mu\nu}[\eta; \pi, \xi]  \ll M_{pl}^2 {\cal E} h_{\mu\nu} $, where $
T_\text{scalar}^{\mu\nu}[g; \pi, \xi]=\frac{2}{\sqrt{-g}} \frac{\delta }{\delta g_{\mu\nu}}\int d^4 x \sqrt{-g}~{\cal L}_\text{scalar}
$ and ${\cal L}_\text{scalar}[g; \pi, \xi]$ is constructed out of the covariant completion of $\hat{ \cal  L}_{\pi, \xi}={ \cal  L}_{\pi, \xi}-3M_{pl}^2 \pi \pd^2 \pi$ , as described in \cite{otherpaper}.  On maximally symmetric backgrounds, this condition schematically corresponds to
\be
T^{\mu\nu}_{scalar}\sim x^2 \sum_{0 \leqslant m+n \leqslant 4} {\cal O}(1)(\alpha_{m,n}-3M_{pl}^2 \delta_m^1 \delta_n^0 )k_\pi^{m+1} k_\xi^n  +{\cal O}(1) \beta_{m,n} k_\pi^{m} k_\xi^{n+1} \ll M_{pl}^2 H^2 \label{backreaction}
\ee

\section{Self acceleration} \label{sec:selfacc}
We begin our analysis with self accelerating vacua. A self accelerating vacuum is one that accelerates even in the absence of any sources for the physical fields, $\tilde h_{\mu\nu}$ and $\pi$. There is some ambiguity as to what is actually meant by this if tadpole terms are present. The point is that at the level of the graviton equations of motion, the source corresponds to the vacuum energy, $\sigma$. However at the level of the scalar equations of motion, we note that  the tadpole term, $\int d^4 x ~\alpha_{0, 0} \pi$, has the effect of renormalising the vacuum energy seen by the $\pi$ field, $\sigma \to \sigma +\alpha_{0, 0}$. To avoid ``cheating", we set the bare vacuum energy $\sigma=0$, and require the $\pi$-tadpole term to vanish, $\alpha_{0, 0}=0$. This is in line with our assumptions at the end of the previous section. It also guarantees that Minkowksi space is a solution for the physical metric since the field equations can be solved by $\tilde h_{\mu\nu}=0, \pi =0$. Note that this Minkowski solution need not be stable. Indeed, our interest is in stable de Sitter solutions. Given the constraints  $\sigma=0, ~\alpha_{0, 0}=0$, any de Sitter solutions are necessarily self accelerating.

Now, since $\sigma=0$, the corresponding GR solution is {\it always} Minkowski, so for {\it any} maximally symmetric solution, we have  $\tilde h_{\mu\nu}=0$. It is non-trivial solutions for the scalar $\pi$ that enable self-acceleration.  Recall that $\bar \pi(x)=-\frac{1}{4}k_\pi x_\mu x^\mu,~\bar \xi(x)=-\frac{1}{4}k_\xi x_\mu x^\mu$ , so for self acceleration we require a solution with $k_\pi \sim H_0^2$, where $H_0$ is the current Hubble scale. For such a solution to be stable, it must correspond to a minimum of the action polynomial. It is easy to build a suitable action polynomial as we will now demonstrate.

Let us assume that a stable self-accelerating vacuum exists, with $k_\pi=H_0^2, ~k_\xi=\zeta H_0^2$, where $\zeta$  is real and can be either positive or negative. We can build the action polynomial about this solution by performing a Taylor expansion:
\be \label{selfaccL}
L(k_\pi, k_\xi)=-\sum_{1 \leqslant m+n\leqslant 4}(\alpha'_{m,n}(k_\pi-H_0^2)+\beta'_{m, n}( k_\xi-\zeta H_0^2)) \left(-\frac{1}{2}\right)^{m+n} \frac{4!}{(4-m-n)!}(k_\pi-H_0^2)^m (k_\xi-\zeta H_0^2)^n  \,.
\ee
Note the lower value of $m+n$ in the sum above; by starting the sum at $m+n=1$ we ensure that $k_\pi=H_0^2, ~k_\xi=\zeta H_0^2$ is indeed a solution. One can straightforwardly check using Eqs. \ref{apr}, \ref{bpr} and \ref{relpr}  that the coefficients are indeed the same $\alpha_{m,n}'$ and $\beta'_{m,n}$ defined in the previous section.
To avoid a ghost on this solution, we require the Hessian of $L$  at $k_\pi=H_0^2, ~k_\xi=\zeta H_0^2$, to have non-negative eigenvalues, or equivalently,
\be  \label{nogh}
\alpha'_{10} \geqslant 0, \qquad \beta'_{01} \geqslant 0, \qquad \alpha'_{10} \beta'_{01} \geqslant \left(\frac{\alpha'_{01}+\beta'_{10}}{2}\right)^2
\ee
As this corresponds to self-acceleration, we  have set $\sigma=0$, and further  require the $\pi$-tadpole to vanish, 
\be 
0=\alpha_{0,0}=-\frac{\pd L}{\pd k_\pi}\Bigg |_{k_\pi=k_\xi=0}=\sum_{1 \leqslant m+n \leqslant 4} \frac{4!}{(4-m-n)!}  \left(\frac{H_0^2}{2}\right)^{m+n} \zeta^n \left[ (m+1)\alpha'_{m,n}+m\beta'_{m,n} \zeta\right] \label{tadpole}
\ee
The right hand side of Eq. \ref{tadpole} is a quartic polynomial in $\zeta$.  We can certainly choose the coefficients so that their exists at least one real root, thereby guaranteeing self acceleration.  Perhaps the simplest example of a theory with a stable self-accelerating vacuum has the solution $\zeta=0$, so that
\be
\sum_{1 \leqslant m \leqslant 4} \frac{4!}{(4-m)!}  \left(\frac{H_0^2}{2}\right)^{m}  (m+1)\alpha'_{m,0}=0 \label{tadpole1}
\ee
We can regard this equation as fixing $\alpha'_{4, 0}$, having already chosen suitable values for  $\alpha'_{1, 0}$,  $\alpha'_{2, 0}$, and  $\alpha'_{3, 0}$.  Indeed, let us assume that 
$$
\alpha'_{m,n}=\frac{\mu^2}{H_0^{2(m+n-1)}} u_{m,n} \qquad \beta'_{m,n}=\frac{\mu^2}{H_0^{2(m+n-1)}} v_{m,n}
$$
where $u_{m,n}$ and $v_{m,n}$ are dimensionless numbers of order one, unless otherwise stated. 
The condition  (\ref{tadpole1}) for vanishing $\pi$ tadpole now takes the simpler form   
\be
\sum_{1 \leqslant m \leqslant 4} \frac{4!}{(4-m)!}  \left(\frac{1}{2}\right)^{m}  (m+1)u_{m,0}=0 \label{tadpole2}
\ee
which we  think of as fixing $u_{4, 0}$, having already chosen  $u_{1, 0}$,  $u_{2, 0}$, and  $u_{3, 0}$. Furthermore, in order to guarantee stability, recall that we must  choose  $u_{1, 0}$,  $v_{0, 1}$ and  $u_{0, 1}+v_{1,0}$ so that Eq. \ref{nogh} holds. 

Note that the overall scale $\mu$ will typically  control the strength of the linearised coupling to matter, whereas the Hubble  scale, $H_0$,  controls  the higher order $\pd \pd \pi$, $\pd \pd \xi$  interaction terms. The action, which can be derived exactly from the action polynomial by  computing the bare coefficients $\alpha_{m,n}$ and $\beta_{m,n}$,  will now take the schematic form
\begin{multline}
S_{scalar}[\pi, \xi; \Psi_n]=\int d^4 x ~\mu^2\left[ \pd \pi \pd \pi  F_1\left(\frac{\pd \pd \pi}{H_0^2}, \frac{\pd \pd \xi}{H_0^2}\right) +2\pd \pi \pd \xi  F_2\left(\frac{\pd \pd \pi}{H_0^2}, \frac{\pd \pd \xi}{H_0^2}\right) +\pd \xi \pd \xi  F_3\left(\frac{\pd \pd \pi}{H_0^2}, \frac{\pd \pd \xi}{H_0^2}\right) \right] \\
+(\textrm{possible $\xi$ tadpole})+\pi T \qquad\qquad
\end{multline}

Let us now  consider the issue of backreaction.  In order to trust our galileon description, we require that the scalar fields do not backreact too much on the geometry.  For the self-accelerating scenario we have just described, the condition (\ref{backreaction}) amounts to $x^2 \mu^2 H_0^4 \ll M_{pl}^2 H_0^2$. Since we restrict attention to subhorizon distances and sub Hubble times, it follows that $\mu \lesssim M_{pl}$, which  is consistent with similar conclusions drawn in the case of the single galileon field \cite{gal}.

\subsection{Spherically symmetric solutions and the Vainshtein effect} \label{sec:selfacc:vain}
One of the features of  self accelerating vacua is that they require ${\cal O}(1)$ modifications of General Relativity on cosmological scales. If such large deviations from GR still occur at much shorter scales, within the solar system, then it is clear that we will not be able to pass local gravity tests  \cite{gravtest}. We need a mechanism to suppress the scalars in the solar system. We know of two: the chameleon mechanism \cite{cham}, which makes use of a non-linear coupling to matter, and the Vainshtein mechanism which makes use of  non-linear self interactions \cite{vainshtein}. Since we have assumed a linear coupling to matter, we are forced to appeal to the latter to help screen the scalar degrees of freedom at short scales.

 Although the Vainshtein mechanism is intimately related to strong coupling in modified gravity theories \cite{dvalithm}, it is a completely classical effect that is still not fully understood. The basic idea is that linearised perturbation theory around a heavy source breaks down at some ``Vainshtein" scale,  $r_V$, below which non-linearities become important, helping to suppress fluctuations in the scalar field relative to the graviton fluctuations. Typically a heavy source like the Sun is treated as a point particle with a definite Vainshtein radius. This is, of course, an over simplification. The Sun is an extended object, made up of many point particles each with their own Vainshtein radii. Is the Vainshtein radius of an extended body the same as the Vainshtein radius of a point particle with the same mass located at the centre of mass? The answer is not known, although given the role of non-linearities in the Vainshtein mechanism, one might expect the answer to be negative. A detailed study is beyond the scope of this paper, although see \cite{testingv, ep} for some work along these lines. 

Having recognised some of the possible pit falls with  this mechanism, we cautiously proceed in the usual way, treating the Sun as a point particle of mass $M_\odot \sim 10^{39} M_{pl}$. We will be interested in spherically symmetric profiles for the scalar field that asymptote to the self-accelerating vacuum, looking to establish the scale at which the linearised theory breaks down, and checking to see if the scalars are indeed screened below that scale. To this end, we will consider spherically symmetric fluctuations about a generic vacuum solution, which we will ultimately take to be the self-accelerating solution just described.  Consistent with our model of the Sun, these fluctuations will be due to a point source of mass $M$ located at the origin. This means we have
$$
\pi = \bar \pi (x)+\pi_s(r), \qquad \xi = \bar \xi (x)+\xi_s(r)
$$
where the background fields, $\bar \pi~\bar \xi$ are given by Eq. \ref{bg}, and the fluctuations $\pi_s(r), ~\xi_s(r)$ satisfy the field equations (\ref{eomxids}) with a source  $\delta T^{\mu}_{\nu}=M\delta^3({\bf r})~\text{diag}(-1, {\bf 0})$. By explicit calculation, we have
\begin{align}
\mathcal{E}_{1,0}&=\frac1{r^2}\frac{\ud}{\ud r}(r^2\pi_s')& & \nn 
\mathcal{E}_{2,0}&=\frac2{r^2}\frac{\ud}{\ud r}(r{\pi_s'}^2) & \mathcal{E}_{1,1}&=\frac2{r^2}\frac{\ud}{\ud r}(r\pi_s'\xi_s') \nn
\mathcal{E}_{3,0}&=\frac2{r^2}\frac{\ud}{\ud r}({\pi_s'}^3) &
\mathcal{E}_{2,1}&=\frac2{r^2}\frac{\ud}{\ud r}(r{\pi_s'}^2\xi_s') \nn
\mathcal{E}_{4,0}&=\mathcal{E}_{3,1}=\mathcal{E}_{2,2}=0, &  &
\end{align}
and for $m<n$ we take  $\mathcal{E}_{m,n}=\mathcal{E}_{n,m}|_{\pi\leftrightarrow\xi}$. Note that $\pi_s'=\pd_r \pi_s$ and $ \xi_s'=\pd_r \xi_s$.  Defining $y=\pi_s'/r$ and $z=\xi_s'/r$, the equations of motion become
\bea \label{ssceom1}
\frac1{r^2}\frac{\ud}{\ud r}[r^3{\cal A}(y,z)]
&=& M \delta^3(\mathbf{r})\\
\label{ssceom2}\frac1{r^2}\frac{\ud}{\ud r}[r^3{\cal B}(y,z)]
&=& 0,
\eea
where
\bea
{\cal A}(y,\!z) &=&  f_1^a+2(f_2^a+f_3^a)\label{sph1},\qquad
{\cal B}(y,\!z) =  f_1^b+2(f_2^b+f_3^b)\label{sph2}\\
f_n^a&=&\sum_{i=0}^n a'_{i, n-i} y^i z^{n-i}, \qquad  f_n^b=\sum_{i=0}^n b'_{i, n-i} y^i z^{n-i}\label{eq:fdef}
\eea
By integrating (\ref{ssceom1}) and (\ref{ssceom2}) over a sphere, we can recast the equations of motion  as two algebraic equations
\bea \label{ssceom3}
{\cal A}(y,z)&=&\frac{M}{4\pi r^3}\\
\label{ssceom4} {\cal B}(y,z)&=&0.
\eea
In the linearised regime, we have
\be
\left(\begin{array}{cc}a'_{1, 0} &  a'_{0,1} \\ b'_{1, 0} & b'_{0, 1}\end{array}\right) \left(\begin{array}{c} y_{lin} \\ z_{lin} \end{array}\right) =\left(\begin{array}{c} M/4\pi r^3 \\ 0 \end{array}\right)
\ee
and so\footnote{Recall from our previous paper \cite{otherpaper} that $\left(\begin{array}{cc}a'_{1, 0} &  a'_{0,1} \\ b'_{1, 0} & b'_{0, 1}\end{array}\right)= \frac12 Hess(L)$.}
\be 
y_{lin}=\frac{4b'_{0, 1}}{\det [Hess(L)]} \frac{M}{4\pi r^3}, \qquad z_{lin}=-\frac{4b'_{1, 0}}{\det [Hess(L)]} \frac{M}{4\pi r^3}
\ee
We can easily solve these differential equations to arrive at the linearised solution for the scalars
\be
\pi^{(lin)}_s(r) =-\frac{4b'_{0, 1}}{\det [Hess(L)]} \frac{M}{4\pi r}, \qquad \xi^{(lin)}_s(r) =\frac{4b'_{1, 0}}{\det [Hess(L)]} \frac{M}{4\pi r}
\ee
To discuss the Vainshtein effect in detail, we now focus our attention on the self accelerating scenario described previously. In general we have $b'_{0, 1} \sim  b'_{1,0} \sim \mu^2$,  and $\det [Hess(L)] \sim \mu^4$, which means $|\pi_s^{(lin)} |\sim | \xi_s^{(lin)}| \sim  \frac{M}{\mu^2 }\frac{1}{ r}$. Assuming the linearised approximation is valid, how much does the $\pi$  mode cause deviations from GR? To answer this consider the graviton fluctuation, corresponding to the standard Newtonian potential, $(\tilde h_{\mu\nu})_N \sim \frac{M}{M_{pl}^2} \frac{1}{r}$, and compare it to the linearised field $\pi^{(lin)}_s$, 
\be
\left|\frac{\pi_s^{(lin)}}{(\tilde h_{\mu\nu})_N}\right| \sim \frac{M_{pl}^2}{\mu^2} \gtrsim 1
\ee
 where the inequality follows from the absence of backreaction onto the vacuum. On solar system scales, deviations from GR are constrained to within one part in $10^5$ \cite{gravtest}. Since  deviations from GR are at least of order one  in the linearised theory, and one has to resort to the Vainshtein effect to pass solar system tests, as expected. 

The Vainshtein effect kicks in when the linearised theory breaks down. Taking the generic scales this occurs when $y_{lin} \sim z_{lin} \sim H_0^2$. It follows that the Vainshtein radius is given by
\be
r_V \sim \left(\frac{M}{\mu^2 H_0^2}\right)^{1/3} \sim \left(\frac{M_{pl}}{\mu}\right)^{2/3} \left(\frac{M}{M_{pl}^2 H_0^2}\right)^{1/3} \gtrsim  10^3 \left(\frac{10^{-9}M}{M_{pl}^2 H_0^2}\right)^{1/3} 
\ee
Ideally we would like the Vainshtein radius around the Sun to exceed the size of the solar system. The solar system extends out to the Oort cloud,  which is of the order $10^{16}$ m from the Sun.   Given that the Sun has mass $M_\odot \sim 10^{39} M_{pl}$ it turns  out  we can write $r_\textrm{oort} \sim \left(\frac{10^{-9}M_\odot}{M_{pl}^2 H_0^2}\right)^{1/3} \sim 10^{16}$ m, and so the Vainshtein radius is at least three orders of magnitude larger than the solar system.

It is not enough to prove that linearised theory breaks down below the Vainshtein radius, we also need to establish whether or not the Vainshtein mechanism actually takes place. In other words, does the $\pi$ mode get screened at distances $r < r_V$? Generically, at short distances, the cubic terms in  Eqs. \ref{sph1} and \ref{sph2} will dominate, and we have 
\be
y \sim z \sim \frac{1}{r} \left(\frac{M H_0^4}{\mu^2}\right)^{1/3} \label{ynonlin}
\ee
It follows that  for $r< r_V$, we have $|\pi_s| \sim |\xi_s| \sim  \left(\frac{MH_0^4}{\mu^2}\right)^{1/3} r$, and so $\left|\frac{\pi_s}{(\tilde h_{\mu\nu})_N}\right| \sim (r/r_0)^2$, where
\be 
r_0 \sim   \left(\frac{\mu}{M_{pl}}\right)^{1/3}\left(\frac{M}{M_{pl}^2 H_0^2}\right)^{1/3} \sim 10^3  \left(\frac{\mu}{M_{pl}}\right)^{1/3} \left(\frac{10^{-9} M}{M_{pl}^2 H_0^2}\right)^{1/3} 
\ee
Now let us evaluate $\left|\frac{\pi_s}{(\tilde h_{\mu\nu})_N}\right|$ in the solar system, up to the maximum distance $r_\text{oort} \sim\left( \frac{10^{-9} M_\odot}{M_{pl}^2 H_0^2} \right)^{1/3}\sim 10^{16}$ m. It follows that within the solar system
\be
\left|\frac{\pi_s}{(\tilde h_{\mu\nu})_N}\right| \lesssim (r_\text{oort}/r_0)^2 \sim 10^{-6} \left(\frac{M_{pl}}{\mu}\right)^{2/3}
\ee
Since deviations from GR should not exceed more that one part in $10^5$ in this region \cite{gravtest}, let us take $\mu \sim M_{pl}$, so that we  safely pass tests of GR in the solar system without running into problems with backreaction on the vacuum. 

Of course we should not only be worrying about backreaction on the vacuum, but backreaction on the spherically symmetric solution itself.  For distances greater than the Vainshtein scale, the linearised theory holds and so the question of backreaction is dominated by the background, where it is known not to be a problem when $\mu \sim M_{pl}$. Below the Vainshtein scale, things are a little more complicated. By taking $\pi\sim\bar\pi+\pi_s\sim H_0^2+r^2y$, $\del\del\pi\sim H_0^2+y$, with $y\sim H_0^2\frac{r_V}{r}$, along with the analogous results for $\xi$,   we can show that schematically, $M_{pl}^2{\cal E} h_{\mu\nu} \sim M/r^3$ and 
\be
T^{\mu\nu}_\textrm{scalar} \sim r^2 M_{pl}^2 H_0^4 \sum_{1 \leqslant m+n \leqslant 4} {\cal O}(1)\left( \frac{r_V}{r}\right)^{m+n+1}  
\ee
 It follows immediately that 
 \be
 \left|\frac{T^{\mu\nu}_\textrm{scalar}}{M_{pl}^2 {\cal E} h_{\mu\nu}}\right| \sim (r_V H_0)^2 \sum_{1 \leqslant m+n \leqslant 4} {\cal O}(1)\left( \frac{r_V}{r}\right)^{m+n-4} \ll 1
\ee
for $r<r_V$. Thus backreaction of the scalars is not an issue, even below the Vainshtein scale, and we can trust our solution all the way down to the Schwarschild radius.

Let us finish this section with a word on strong coupling. Strong coupling is inevitably linked to the Vainshtein effect \cite{dvalithm} on a given background. Therefore, since the Vainshtein mechanism takes place, we expect quantum fluctuations about the background vacuum to become strong coupled at reasonably large scales. To elucidate this let us consider the effective action (\ref{pertact}) describing excitations about non-trivial vacua. Focussing on the scalar sector, we first diagonalise the kinetic term, by performing a linear map of ${\cal O}(1)$ on the scalars, $(\pi, \xi) \to (\pi_*, \xi_*)$. Schematically, the scalar action now takes the form
\begin{multline}
S_{scalar}[\pi_*, \xi_*; \Psi_n]=\int d^4 x~  -\frac{\lambda_1}{2} (\pd \pi_*)^2 -\frac{\lambda_2}{2} ( \pd \xi_*)^2+( {\cal O}(1)\pi_*+ {\cal O}(1)\xi_*)\delta T\\
+\mu^2 \sum_{2 \leqslant m+n \leqslant 4} \left( {\cal O}(1)\pd \pi_* \pd \pi_*+{\cal O}(1) \pd \pi_* \pd \xi_*+{\cal O}(1) \pd \pi_* \pd \xi_*\right)\left(\frac{ \pd \pd  \pi_*}{H_0^2}\right)^{m-1}\left(\frac{ \pd \pd \xi_*}{H_0^2}\right)^{n-1}
\end{multline}
where $\lambda_1$ and $\lambda_2$ are the eigenvalues of the Hessian, $Hess(L)$. Both eigenvalues are positive in a ghost free theory, and, for self accelerating vacua,  of order $\mu^2$.  Now let us canonically normalise the scalar fields by defining $\hat \pi=\sqrt{\lambda_1} \pi_*$ and $\hat \xi=\sqrt{\lambda_2} \xi_*$, so that
\begin{multline}
S_{scalar}[\hat \pi, \hat \xi; \Psi_n]=\int d^4 x~  -\frac{1}{2} (\pd  \hat  \pi)^2 -\frac{1}{2} ( \pd  \hat  \xi)^2+\frac{1}{\mu}( {\cal O}(1)\hat \pi+ {\cal O}(1)\hat \xi)\delta T\\
+\sum_{2 \leqslant m+n \leqslant 4} \left( {\cal O}(1)\pd \hat \pi \pd\hat \pi+{\cal O}(1) \pd  \hat \pi\pd  \hat \xi+{\cal O}(1) \pd \hat  \pi \pd \hat  \xi \right)\left(\frac{ \pd \pd  \hat  \pi}{\mu H_0^2}\right)^{m-1}\left(\frac{ \pd \pd  \hat \xi }{\mu H_0^2}\right)^{n-1}
\end{multline}
Here we see explicitly how $\mu$ controls the strength of the scalar coupling to matter. Given that this is Planckian, the scalars couple to matter with gravitational strength. Furthermore, it is clear that the interactions become strongly coupled at a scale $\Lambda_0 \sim (\mu H_0^2)^{1/3}$. For $\mu \sim M_{pl}$, this means that tree level interactions are only valid down to distances of the order $\Lambda_0^{-1} \sim 1000$ km, below which loop effects are important. This might lead one to question the validity of our classical description. However, it is important to realise that these scales correspond to modes propagating on the vacuum. Our real concern lies with quantum fluctuations propagating on the non-trivial solution around the heavy classical source \cite{nicolis}. We will study fluctuations about the spherically symmetric solutions in the next section. 

\subsection{Fluctuations about spherical symmetry}
In the analysis of the single galileon \cite{gal},  promising solutions were found to exist, exhibiting self-acceleration without ghosts and with a suitable Vainshtein mechanism occurring within the solar system. However, fluctuations about the corresponding spherically symmetric profiles revealed a number of insurmountable problems. For example, at large distances, radial modes were found to propagate superluminally, leading to worries about causality. In contrast, at shorter distances, angular modes were found to propagate extremely slowly, so much so that the earth's motion through the solar galileon field would result in excessive emission of Cerenkov radiation, raising serious doubts about the validity of the  static approximation. A third problem pertained to the domain of validity of the classical theory set by the scale of strong coupling on the spherically symmetric background. Whilst it was hoped that strong coupling would occur at higher energies than in the corresponding vacuum, the opposite  was generically true.  None of these pathologies could be eliminated in an entirely acceptable manner. The related problems of Cerenkov radiation and low scale of strong coupling could be alleviated by suppressing quartic and quintic interactions. However, to avoid a ghost in this scenario one needed to introduce a tadpole, which is undesirable as it amounts to a renormalisation of the vacuum energy to non-zero values. To further avoid issues with superluminality, one would have to do away with all higher order interactions and so  abandon all hope of Vainshtein screening in the solar system. In summary, it seemed that a fully consistent self-accelerating scenario could not be found.

In this section we will show that {\it  all} of these pathologies can be {\it simultaneously} avoided in the case of two galileons. To see this, let us consider small fluctuations about the spherically symmetric solutions discussed in the previous section,
\be
\pi=\pi_s+\phi,\qquad\xi=\xi_s+\psi\label{eq:sphPert}
\ee
Although we will ultimately be interested in the scale of higher order interactions, for the moment we will focus on the leading order theory given by the quadratic Lagrangian
\be
\mathcal{L}_{\phi,\psi}
=\frac12\pd_t{\Phi}\cdot \mathcal{K}\pd_t{\Phi}-\frac12 \pd_r{\Phi}\cdot\mathcal{U}{\pd_r\Phi }-\frac12 \pd_\Omega{\Phi}\cdot \mathcal{V} \pd_\Omega{\Phi}\label{eq:Keq}
\ee
where the angular derivative $\pd_\Omega=\mathbf{\hat{e}}_\theta \frac1r \pd_\theta +\mathbf{\hat{e}}_\varphi \frac1{r \mathrm{sin}\theta} \pd_\varphi$, and dot products between angular derivatives are understood.  Note that the fluctuations form a 2 component vector $\Phi=\left(\begin{array} {c}\phi \\ \psi\end{array}\right)$ and so the kinetic mixing matrices take the form
\be 
{\cal K} =\left[ \begin {array}{cc} K_{\phi\phi} & K_{\phi\psi}\\\noalign{\medskip}
K_{\phi \psi} & K_{\psi \psi}\end {array} \right], \qquad {\cal U}=\left[ \begin {array}{cc} U_{\phi \phi} & U_{\phi \psi}\\\noalign{\medskip}
U_{\phi \psi} & U_{\psi \psi}\end {array} \right], \qquad {\cal V}=\left[ \begin {array}{cc} V_{\phi \phi} & V_{\phi \psi} \\\noalign{\medskip}
V_{\phi \psi} & V_{\psi \psi} \end {array} \right]
\ee
These matrices can be calculated explicitly by, for example, working out the equations of motion coming from (\ref{eq:Keq}), then comparing this with the linearised equations of motion of the system. By using (\ref{eq:sphPert}) and $y=\pi_s'/r$, $z=\xi_s'/r$, the results are most elegantly expressed using (\ref{eq:fdef}) as follows
\be \label{KV}
{\cal K}={\cal J} +\frac{r}{3} \pd_r {\cal J}, \qquad {\cal V}={\cal U}+\frac{r}{2} \pd_r{\cal U}
\ee
where
\be \label{UJ}
{\cal U}=\Sigma_1+2\Sigma_2 +2 \Sigma_3, \qquad {\cal J}=\Sigma_1+3\Sigma_2+6\Sigma_3+6\Sigma_4, 
\ee
and
\be \label{sigma}
\Sigma_n=\left[ \begin {array}{cc} \pd_y f_n^a & \pd_y f_n^b \\\noalign{\medskip} \pd_z f_n^a & \pd_z f_n^b\end {array} \right],
\ee
where $f_n^a$ and $f_n^b$ are given by Eq. \ref{eq:fdef}.

\subsubsection*{Ghosts, tachyons and superluminality}
Now the resulting equations of motion describe the linearised perturbation theory, and take following form
\be
-{\cal K} \pd_t^2 \Phi +\frac1{r^2}\pd_r[ r^2 {\cal U} \pd_r \Phi ]+{\cal V} \pd^2_\Omega  \Phi=0
\ee
This is all we need to study a number of issues, including the speed of mode propagation, as well as possible instabilities arising from ghosts and tachyons. Indeed, we can read off the ``no ghost condition" immediately. We simply require that ${\cal K}$ has non-negative eigenvalues, or equivalently
$$K_{\phi \phi} \geq 0, \qquad K_{\psi \psi} \geq 0, \qquad K_{\phi \phi}K_{\psi \psi} \geq K_{\phi \psi}^2$$
To address the other issues we need to derive the dispersion relations for the normal modes of the system. This is a little involved since these do not  correspond to  $\phi$ and $\psi$ if the cross terms $K_{\phi \psi}, ~U_{\phi \psi}, V_{\phi \psi}$ are non-zero.  To derive the dispersion relations for the normal modes,  we first  assume that the background changes slowly with radius compared to the fluctuations, so that we can treat it as roughly constant. The equation of motion becomes $-{\cal K}\pd_t^2 \Phi +{\cal U} \pd_r^2 \Phi +{\cal V} \pd^2_\Omega  \Phi \approx 0$, where the  radial variation in the mixing matrices is being neglected.  We now take Fourier transforms
\be \label{disp}
\left[{\cal K} \omega^2  -{\cal U} p_r^2  -{\cal V} p^2_\Omega\right] \tilde  \Phi(\omega, p_r, p_\Omega) \approx 0
\ee
where $p_r$ and $p_\Omega$ are the momenta along the radial and angular directions respectively. It follows that the dispersion relations are given implicitly by $\det \left[{\cal K} \omega^2  -{\cal U} p_r^2  -{\cal V} p^2_\Omega\right] =0$. Now it is convenient to write $p_r=p \cos q, p_\Omega=p\sin q$, so that the dispersion relations for the two eigenmodes can be written as $\omega^2=c_\pm^2(q) p^2$. The sound speeds along the $q$ direction are given by
\be
c_\pm^2(q) =\frac12\left[Tr {\cal M}(q) \pm \sqrt{(Tr {\cal M}(q))^2-4 \det {\cal M}(q)} \right]
\ee
where ${\cal M}(q)={\cal K}^{-1}{\cal U} \cos^2 q +{\cal K}^{-1}{\cal V}\sin^2 q$.  We now impose the condition $0 \leq c_\pm^2(q) \leq 1$, which should be valid for all $q$. The lower bound prevents tachyonic instability, whereas the upper bound prevents superluminal mode propagation. We can make these conditions more explicit. The ``no tachyon"  condition, $c^2_\pm  \geq  0$ is equivalent to requiring that ${\cal M}$ has non-negative eigenvalues\footnote{In fact, the eigenvalues of ${\cal M}$ are just $c_\pm^2$.}. In contrast, the ``no superluminality" condition, $c_\pm^2(q) \leq 1$ is equivalent to requiring that ${\cal M}-I$ has non-{\it positive} eigenvalues. Writing ${\cal M}=\left[ \begin {array}{cc} M_{\phi \phi} & M_{\phi \psi}\\\noalign{\medskip}
M_{\psi \phi} & M_{\psi \psi}\end {array} \right]$, these requirements  correspond to
\bea
\textrm{``no tachyon"} && M_{\phi \phi} +M_{\psi \psi} \geq 0, \qquad M_{\phi \phi}M_{\psi \psi} \geq M_{\phi \psi}M_{\psi \phi}   \\
 \textrm{``no superluminality"} && M_{\phi \phi} + M_{\psi \psi} \leq 2, \qquad (1-M_{\phi \phi})(1-M_{\psi \psi}) \geq M_{\phi \psi}M_{\psi \phi}
\eea

\subsubsection*{Large distance behaviour}

Let us now study these conditions more closely. We begin with an analysis of the behaviour at large distances, $r \gg r_V$. In the single galileon case, the radial modes become superluminal at this scale \cite{gal}.  To see whether the same thing happens here we calculate the matrices perturbatively. First, we consider the equations of motion Eqs.~\ref{ssceom3} and \ref{ssceom4} order by order, using the ansatz  $y=y^{(l)}+y^{(nl)}+...$ and $y=z^{(l)}+z^{(nl)}+...$, where ``l" labels  the ``leading order" contribution, ``nl"  labels ``next to leading order", and so on. Note that we suppress the evaluation of the expressions $|_{r\gg r_V}$ except for the final results. Generically, we expect
\be
y^{(l)},z^{(l)}\sim \frac{1}{r^3}, \qquad  y^{(nl)},z^{(nl)}\sim \frac{1}{r^6} , \qquad \ldots
\ee
which lead to the relations $\pd_r y^{(l)}\sim -3 y^{(l)} /r$, $\pd_r y^{(nl)}\sim -6y^{(nl)}/r$, ... (and similar relations for $z$). We also expand $\Sigma_n=\Sigma^{(l)}_n+\Sigma^{(nl)}_n+...$ (except for $\Sigma_1$, which is constant), and it follows that
\be
r\pd_r \Sigma^{(l)}_2\approx-3\Sigma^{(l)}_2, \qquad  r\pd_r \Sigma^{(nl)}_2\approx-6\Sigma^{(nl)}_2 ,\qquad
r\pd_r \Sigma^{(l)}_3\approx-6\Sigma^{(l)}_3, \qquad \ldots
\ee
Notice that the next to leading term $\Sigma^{(nl)}_2$ is of the same order of the leading term $\Sigma^{(l)}_3$. Plugging this into our formulae Eqs.~\ref{KV}, \ref{UJ} and \ref{sigma},  we find
\be
{\cal K} \approx \Sigma_1-3 \Sigma^{(nl)}_2-6\Sigma^{(l)}_3, \qquad {\cal U}\approx\Sigma_1 +2\Sigma_2^{(l)} +2(\Sigma^{(nl)}_2+ \Sigma^{(l)}_3), \qquad {\cal V}\approx \Sigma_1 -\Sigma_2^{(l)}-4 (\Sigma^{(nl)}_2 + \Sigma^{(l)}_3)
\ee
where we have ignored $O(\frac{1}{r^9})$ terms. To compute ${\cal M}$ we need knowledge of ${\cal K}^{-1}$, which we also calculate perturbatively
\be
{\cal K}^{-1}=\Sigma^{-1}_1 +6\Sigma^{-1}_1 ( \Sigma_2^{(nl)}+\Sigma_3^{(l)} )\Sigma^{-1}_1+\ldots
\ee
and so
\be
{\cal M}=I+(3\cos^2q-1) \Sigma_1^{-1} \Sigma_2 ^{(l)}+(6\cos^2q-1) \Sigma_1^{-1} \Sigma_2^{(nl)} + (6\cos^2q+2) \Sigma_1^{-1}  \Sigma_3^{(l)}+\ldots
\ee
Now by studying the leading order contribution, it is immediately clear that,  generically, the eigenvalues of ${\cal M}-I$ will change sign at $\cos q =\pm 1/\sqrt{3}$, meaning that propagations along some directions will be superluminal. This can be avoided if we impose the condition $\left.\Sigma^{(l)}_2\right|_{r \gg r_V}=\left.\Sigma^{(nl)}_2\right|_{r \gg r_V}=0$ and require
$\left.\Sigma_1^{-1} \Sigma_3^{(l)}\right|_{r\gg r_V}$ has negative eigenvalues. This guarantees that $M-I$ has negative eigenvalues to leading order, ensuring that all modes propagate subluminally. We also see that there is never an issue with tachyonic instability at large distances  since the sound speeds are always close to unity. Furthermore, as long as the vacuum is ghost free, then we have that ${\cal K}$ has non-negative eigenvalues to leading order ($\Sigma_1$ is non-negative definite), and therefore there are no ghosts and the condition that $\left.\Sigma_1^{-1} \Sigma_3^{(l)}\right|_{r\gg r_V}$ is negative definite reduces to that $\left.\Sigma_3^{(l)}\right|_{r\gg r_V}$ is negative definite.

\subsubsection*{Short distance behaviour}

We now turn our attention to shorter distances, below the Vainshtein radius, $r \ll r_V$. For the single galileon, short distance modes generically propagate very slowly along the angular direction. In fact, the propagation is so slow that one gets excessive emission of Cerenkov radiation as the earth moves through the solar galileon field, raising serious doubts as to the validity of our static approximation. This problem  can be avoided by eliminating quartic and quintic interactions of the single galileon, although in the absence of a tadpole, this procedure renders the self-accelerating vacuum unstable to ghostly excitations. We will find that such difficulties can be avoided in the case of two galileons, even without having to introduce a tadpole.

We first do the perturbative expansion at short distances: $y=y^{(l)}+y^{(nl)}+...$, $y=z^{(l)}+z^{(nl)}+...$ and $\Sigma_n=\Sigma^{(l)}_n+\Sigma^{(nl)}_n+\Sigma^{(nnl)}_n+...$. Note again that we suppress the evaluation of the expressions $|_{r\ll r_V}$ except for the final results.  By considering the equations of motion order by order, we would generically expect
\be
y^{(l)},z^{(l)}\sim \frac{1}{r}, \qquad  y^{(nl)},z^{(nl)}\sim 1 , \qquad \ldots
\ee
and
\be
\Sigma^{(l)}_4 \sim \frac{1}{r^3}, \qquad \Sigma^{(nl)}_4, \Sigma^{(l)}_3 \sim \frac{1}{r^2}, \qquad  \Sigma^{(nnl)}_4, \Sigma^{(nl)}_3, \Sigma^{(l)}_2 \sim \frac{1}{r}
 \qquad,\ldots
\ee
Similar to the large distance case, we can eliminate $r\pd_r \Sigma_n$ in favour of $\Sigma_n$ and get
\be \label{kuvmat}
{\cal K} \approx  2\Sigma^{(nl)}_4 +4 \Sigma^{(nnl)}_4 + 2\Sigma^{(l)}_3 +4\Sigma^{(nl)}_3 +2\Sigma^{(l)}_2, \qquad
 {\cal U} \approx 2\Sigma^{(l)}_3+2\Sigma^{(nl)}_3+2\Sigma^{(l)}_2, \qquad
 {\cal V} \approx \Sigma^{(nl)}_3 + \Sigma^{(l)}_2
\ee
where we have neglected $O(1)$ terms. As we shall see in next subsection about the strong coupling issue, we can set $\Sigma_4= 0$ (no evaluation at $r\ll r_V$), i.e., set $a'_{m, 4-m}= b'_{m, 4-m}= 0$, to avoid the strong coupling problem. We shall assume this here and the three matrices reduce to
\be
{\cal K} \approx  2\Sigma^{(l)}_3 +4\Sigma^{(nl)}_3 +2\Sigma^{(l)}_2, \qquad
 {\cal U} \approx 2\Sigma^{(l)}_3+2\Sigma^{(nl)}_3+2\Sigma^{(l)}_2, \qquad
 {\cal V} \approx \Sigma^{(nl)}_3 + \Sigma^{(l)}_2
\ee
Now, assuming $\det \Sigma_3^{(l)} \neq 0$, we have ${\cal K}^{-1} \approx \frac12  [\Sigma^{(l)}_3]^{-1}$ and so ${\cal M} \approx \cos^2 q I$. It is clear that  the sound speed along the angular direction, $\cos q=0$, will be very small, and we will once again run into problems with emission of Cerenkov radiation due to the earth's motion through the slowly propagating galileon field. So we enforce the condition $\left.\Sigma_3^{(l)}\right|_{r\ll r_V}=0$. Then we have
\be
{\cal M}=(\cos^2q+1 ) ( 4 \Sigma_3^{(nl)}+2\Sigma_2 ^{(l)} )^{-1}(  \Sigma_3^{(nl)}+\Sigma_2 ^{(l)} )  +  \ldots
\ee
and we should also impose the condition that $\left[(4 \Sigma_3^{(nl)}+2\Sigma_2 ^{(l)} )^{-1}(  \Sigma_3^{(nl)}+\Sigma_2 ^{(l)} )\right]_{r\ll r_V}$ has non-negative eigenvalues.

\subsubsection*{Strong coupling} \label{strongcpl}

We saw in the previous section that quantum fluctuations on the self accelerating {\it vacuum} become strongly coupled below a length scale of around $\Lambda_0^{-1} \sim (\mu H_0^2)^{-1/3} \sim 1000 \textrm{km}$, beyond which one cannot trust the classical description. We are now ready to ask whether or not the corresponding momentum scale can be pushed higher for quantum fluctuations on the spherically symmetric solution. Generically this was not the case for a single galileon, and although the same is true here, we will find examples where there are no issues with strong coupling, or indeed any of the other pathologies we have discussed. For the single galileon, the scale of strong coupling can only be raised at the expense of introducing either a ghost or a tadpole, neither of which is theoretically desirable. Such difficulties are avoided in bi-galileon theory.

Since strong coupling kicks in at short distances, we work in the short distance approximation. To estimate the strong coupling scales above the spherically symmetrical background, we  need to work out the interaction terms in the Lagrangian.  While it is necessary to keep the background coordinates exact, ie, spherical coordinates, it suffices to treat the perturbed coordinates as cartesian and treat the  coefficients as constants, as long as the strong coupling length scale is smaller the radius of the point in question. It is easier to work at  the level of the equation of motion and then promote it to the Lagrangian. Neglecting the details of the contractions, and taking $\pi_s'/r=y$ to imply $\del \del\pi_s\sim y$, along with the analogous results for $\xi$, the full Lagrangian is schematically given by
\begin{multline}
{\cal L}_{\phi,\psi}\approx \mu^2 \frac{w}{H_0^2} \left[ {\cal O}(1)(\pd \phi)^2+{\cal O}(1)(\pd \phi \pd \psi)+{\cal O}(1)(\pd \psi)^2\right] \\
 +\mu^2 H_0^2  \sum_{2 \leq m+n \leq 4} \sum_{2 \leq i+j \leq 4}\left(\frac{w}{H_0^2}\right)^{m+n-(i+j)}\left[{\cal O}(1)\phi+{\cal O}(1)\psi\right]\left(\frac{\pd\pd \phi}{H_0^2}\right)^i\left(\frac{\pd\pd \psi}{H_0^2}\right)^j
\end{multline}
where $w \sim y \sim z $, as given by Eq.  \ref{ynonlin}. Now let us canonically normalise the scalar fields by performing the linear map $\phi=\frac1{\mu} \sqrt{\frac{H_0^2}{w}} \left( {\cal O}(1) \hat \phi+{\cal O}(1) \hat \psi \right)$, $\psi=\frac1{\mu} \sqrt{\frac{H_0^2}{w}} \left( {\cal O}(1) \hat \phi+{\cal O}(1) \hat \psi \right)$, 
\begin{multline}
{\cal L}_{\hat \phi, \hat \psi} \approx {\cal O}(1)(\pd \hat \phi)^2+{\cal O}(1)(\pd \hat \psi)^2\\
+ \sum_{2 \leq m+n \leq 4} \sum_{2 \leq i+j \leq 4} \left(\frac{w}{H_0^2}\right)^{m+n-\frac{3}{2}(i+j)-\frac12} \Lambda_0^{3(1-i-j)}\left[{\cal O}(1)\hat \phi+{\cal O}(1)\hat \psi\right]\left(\pd\pd \hat \phi\right)^i\left(\pd\pd \hat \psi\right)^j
\end{multline}
In (hopefully) obvious notation, these interactions become strongly coupled at a scale
\be
\Lambda_{m+n, i+j} \sim \Lambda_0  \left(\frac{w}{H_0^2}\right)^\frac{2(m+n)-3(i+j)-1} {6(1-i-j)}
\ee
Since $w \gg H_0^2$ at short distances, it follows that the largest interaction comes from the term with $i+j=2$, $m+n=4$, becoming strongly coupled at
\be
\Lambda_{4, 2} \sim \Lambda_0  \left(\frac{w}{H_0^2}\right)^{-1/6},
\ee
Thus strong coupling kicks in at an even lower scale than in the vacuum if such interactions are present. This is clearly undesirable, so we set all terms with $m+n=4$ to vanish, $a'_{m, 4-m}=b'_{m, 4-m}=0$.  This does not come into conflict with any of our previous phenomenological constraints. Things are now much better behaved. The largest interaction comes from the term with $i+j=2$, $m+n=3$, becoming  strongly coupled at a much higher energy scale
\be
\Lambda_{3, 2} \sim \Lambda_0  \left(\frac{w}{H_0^2}\right)^{1/6},
\ee
From Eq.~\ref{ynonlin}, in the solar system we have $w \sim \frac{1}{r} \left(\frac{M_\odot  H_0^4}{\mu^2}\right)^{1/3}$, from which we see that the strong coupling scale runs, and is given by
\be
\Lambda_{sc}(r) \sim 10^{-13/6} \Lambda_0^{5/6} r^{-1/6}
\ee
Our classical description breaks down when $r\Lambda_{sc}(r) \sim 1$, after which quantum corrections become important \cite{nicolis}. This occurs at a critical radius $r_c \sim 10^{-13/5} \Lambda_0^{-1} \sim 2-3 \textrm{~km}$, which is of the order the Schwarzschild radius of the Sun. This is perfectly acceptable from a phenomenological point of view as we don't expect the galileon description to be valid at such a low scale anyway!

\section{Self tuning} \label{sec:selftun}

Now let us turn our attention to {\it self-tuning} vacua. We define a self tuning vacuum as one that is Minkowski,  for {\it any} value of the vacuum energy $\sigma$. Self tuning mechanisms are designed to solve the ``old" cosmological constant problem\footnote{See \cite{relax} for a nice review of the``old" cosmological constant problem.}. This arises because each of the matter fields, $\Psi_n$,  contribute to the overall vacuum energy density, $\sigma=\sum_n \langle \rho_{\Psi_n} \rangle$. Each contribution is found by summing up the zero-point energies, $\frac{1}{2} \hbar \omega_{\vec k}$,  of the  normal modes, up to some cut-off $\Lambda \gg m$, where $m$ is the particle mass. Setting $\hbar=c=1$, this generically gives~\cite{nogo}
\be
 \langle \rho_{\Psi} \rangle =  \int^\Lambda_0 \frac{d^3k}{(2\pi)^3} \frac{1}{2}\sqrt{{\vec k}^2+m^2} \approx \frac{ \Lambda^4}{16 \pi^2}
\ee 
so that  the overall vacuum energy $\sigma \sim \Lambda^4$. We often assume that this cut off is Planckian giving rise to a huge vacuum energy of around $10^{72}~ (\text{GeV})^4$. We can reduce this slightly by allowing for  supersymmetry, so that the cut-off corresponds to the supersymmetry breaking scale, but even then the vacuum energy is  at least  $(\text{TeV})^4$. In any event, in GR, such a huge vacuum energy would cause the vacuum to be highly curved, giving a de Sitter or anti de Sitter geometry with curvature scale $\sim \sqrt{|\sigma|}/M_{pl}$.  In terms of our graviton mode, we have $\tilde h^\textrm{vac}_{\mu\nu}=\frac{\sigma}{6M_{pl}^2} x_\alpha x^\alpha\eta_{\mu\nu}$.

For self tuning to occur, we require the {\it physical} metric to be Minkowski, whatever the choice of $\sigma$. In other words, we have $\bar h_{\mu\nu}=0$, and so $\bar \pi=-\frac{\sigma}{12M_{pl}^2} x_\alpha x^\alpha$ in order to cancel off the contribution from the graviton. It follows that we need $k_\pi=-\frac{\sigma}{3 M_{pl}^2}$ to be a solution to the background field equations,
 $\frac{ \pd L}{\pd k_{\pi}}=\frac{ \pd L}{\pd k_{\xi}}=0$, for {\it any} value of $\sigma$.  
 
It is important to realise that the coefficients $\alpha_{m,n}, ~\beta_{m,n}$ {\it do not depend on $\sigma$}, as these coefficients define the theory and are independent of the source. Having said that, there {\it is} explicit $\sigma$ dependence in the action polynomial (\ref{actpoly})
 \be
 L(k_\pi, k_\xi; \sigma)=4\sigma k_\pi-\sum_{0 \leqslant m+n\leqslant 4}(\alpha_{m,n}k_\pi+\beta_{m, n} k_\xi) \left(-\frac{1}{2}\right)^{m+n} \frac{4!}{(4-m-n)!}k_\pi^m k_\xi^n  \,.
\ee
and so different values of $\sigma$ give different action polynomials. For any given $\sigma$, we require that  the corresponding action polynomial has a {\it minimum}  at $k_\pi=-\frac{\sigma}{3 M_{pl}^2}$, so that our self tuning is stable against ghost-like excitations.
 
Now if the action polynomial $L(k_\pi, k_\xi; \sigma)$ has an extremum at $k_\pi=-\frac{\sigma}{3 M_{pl}^2}$, then it is obvious that the {\it modified} action polynomial
 \be
 \hat L(k_\pi, k_\xi)= L(k_\pi, k_\xi; \sigma)-6M_{pl}^2\left[ \left(k_\pi+\frac{\sigma}{3M_{pl}^2}\right)^2-\frac{\sigma^2}{9 M_{pl}^4}\right]
 \ee
 also has an extremum at this point. However, the interesting thing to note is that $\hat L$ is independent of $\sigma$. It  follows that the modified action polynomial has a {\it continuum} of extrema, $(k_\pi, k_\xi)=\left(\zeta,  f(\zeta)\right)$, parametrized by $\zeta=-\sigma/3M_{pl}^2$.  Now, this is the case if and only if $\hat L$ has the form
 \be
  \hat L(k_\pi, k_\xi)=(k_\xi - f(k_\pi))^2 C(k_\pi, k_\xi)+\text{constant}
  \ee
  where $C(k_\pi, k_\xi)$ is a cubic in $k_\xi$, with $k_\pi$-dependent coefficients. As the constant is irrelevant, we might as well neglect it for simplicity.  We can now think of  $\hat L$ as a quintic  polynomial in $k_\xi$, with a real double root.
  
  To establish whether or not we have ghosts, we need to compute the Hessian of the action polynomial on the solution.  It is easy to check that
  \be
  Hess(L)=Hess(\hat L)+\left(\begin{array}{c c} 12M_{pl}^2 & 0\\ 0 &0 \end{array}\right)
  \ee
It follows that at $(k_\pi, k_\xi)=\left(\zeta,  f(\zeta)\right)$, we have
\be
Hess ( L)|_{\left(\zeta,  f(\zeta)\right)}=\left(\begin{array}{c c} 2f'(\zeta)^2 C(\zeta, f(\zeta)) +12M_{pl}^2 & -2f'(\zeta) C(\zeta, f(\zeta)) \\  -2f'(\zeta) C(\zeta, f(\zeta))  & 2C(\zeta, f(\zeta)) \end{array}\right)
\ee
To avoid ghosts, we require the eigenvalues of this Hessian to be non-negative, so we simply need $C(\zeta, f(\zeta)) \geqslant 0$. Therefore, the action polynomial for a theory admitting stable self-tuning vacua can always be written as
\be \label{selftunL}
L(k_\pi, k_\xi; \sigma)=4\sigma k_\pi+ \sum_{n=2}^5 c_{5-n} (k_\pi) (k_\xi-f(k_\pi))^{n}+6M_{pl}^2 k_\pi^2+\text{constant}
\ee
where $c_3(k_\pi) \geqslant 0$ for  $k_\pi \in \left(-\frac{\sigma_1}{3M_{pl}^2}, -\frac{\sigma_2}{3M_{pl}^2}\right)$. Such a self tuning theory is stable for a range of vacuum energies $\sigma \in (\sigma_2, \sigma_1)$. The function $L(k_\pi, k_\xi)$ ought to be a bivariate polynomial, up to fifth order in the variables $k_\pi$ and $k_\xi$. For generic functions $f, ~c_n$, this will not be the case given the form for $L$ in Eq. \ref{selftunL}. Thus $f$ and $c_n$ ought to be chosen appropriately.

As a simple example, we can choose $f(k_\pi)=M^4/k_\pi$ and $c_3(k_{\pi})=k_{\pi}^2/\mu^2$, with other $c_{5-n}$ vanishing. This gives rise to the self-tuning model:
\be
{\cal L}_{\pi,\xi}= 3M_{pl}^2\pi \Box \pi - \frac{M^4}{\mu^2}  \pi \Box \xi + 
\frac1{3\mu^2} \pi \left[ \Box \pi ( (\Box \xi)^2 - (\pd_\mu \pd_\nu \xi)^2 ) 
- 2 \pd_\mu \pd_\nu \pi \pd^\mu \pd^\nu \!\xi \Box \xi   + 2 \pd_\mu \pd_\nu \pi \pd^\nu \pd^\rho \xi \pd_\rho \pd^\mu \!\xi   \right]
\ee
One might check explicitly that the background equations of motion are solved by $k_{\xi}=M^4/k_{\pi}=-3M^4 M_{pl}^2/ \sigma$, irrespective of the value of vacuum energy $\sigma$. Whilst  we have a ghost on the trivial background ($k_\pi=k_\xi=0$), this is not important  since fluctuations on the  {\it self-tuning} background are ghost free, as one can easily check. The backreaction of the self-tuning background onto the geometry will be negligible if the condition $M^4\lesssim \mu M_{pl} H_0^2$ is satisfied. Vainshtein screening is effective owing to the presence of the higher order galileon terms. Indeed, by performing perturbation analysis on the linear spherically symmetric solution, we can estimate that the Vainshtein radius of the Sun is about $r_V\sim (M_{\odot}M_{pl}M^4/\mu \sigma^2)^{1/3}$. Unfortunately, as we will see in Section \ref{sphback}, one cannot self-tune a large vacuum energy without introducing problems with backreaction.

\subsection{Evading Weinberg's no go theorem}

Let us divert our discussion and consider for a moment our self tuning solution in the context of Weinberg's no go theorem~\cite{nogo}. Weinberg argued, on very general grounds, that no dynamical  adjustment mechanisms could be used to solve the cosmological constant problem.  Let us briefly sketch his proof, adapted to the case at hand. Imagine a system of two scalar fields, $\pi_1, \pi_2$, non-minimally coupled to gravity, described by a general action
\be
S[\pi_i, g_{ab}]=\int d^4 x \sqrt{-g}R+ { L}(\pi_i, g_{\mu\nu}, \pd \pi_i, \pd g_{\mu\nu},  \pd\pd \pi_i, \pd \pd g_{\mu\nu} \ldots)
\ee
We assume that the matter fields, $\Psi_n$, all lie in their ground state, and absorb their contribution to the vacuum energy into the potential for the scalar fields, $V(\pi_1, \pi_2)$. Let us consider a Poincar\'e invariant solution to the field equations, with constant scalars, and ``constant" metric, $g_{\mu\nu}=\eta_{\mu\nu}$. It follows that this solution satisfies the equilibrium condition
\be \label{weinbergeq}
\frac{\pd L}{\pd g_{\mu\nu}}\Big |_{g, \pi=\text{constant}}=0, \qquad \frac{\pd L}{\pd \pi_i}\Big |_{g, \pi=\text{constant}} =0
\ee
For this solution to be ``natural" we demand that the trace of the gravity equation is a linear combination of the scalar equations,
\be \label{lincom}
g_{\mu\nu}\frac{\pd L}{\pd g_{\mu\nu}}=\sum_i \frac{\pd L}{\pd \pi_i}f_i(\pi_i) 
\ee
for all constant fields. This ensures that the trace of the gravity equation vanishes automatically by virtue of the scalar equations of motion\footnote{Eq. \ref{lincom} is a sufficient but not a necessary condition for this to hold.}.  Defining $\chi=\pi_1/2f_1-\pi_2/2f_2$ and $\phi=\pi_1/2f_1+\pi_2/2f_2$, we note that (for constant fields) the Lagrangian $  L$ is invariant under
\be\delta g_{\mu\nu}=\epsilon g_{\mu\nu}, \qquad \delta \chi=0, \qquad \delta \phi=-\epsilon\ee
 It now follows that the Lagrangian must take the form $ L=\sqrt{-\hat g} {\cal L}(\chi, \pd \chi, \pd \phi, \pd \hat g_{\mu\nu}, \pd\pd \ldots)$ where $\hat g_{\mu\nu}=e^{\phi}g_{\mu\nu}$, and so 
 \be
\frac{\pd L}{\pd g_{\mu\nu}}\Big |_{g, \pi=\text{constant}}=\frac{1}{2}g^{\mu\nu}L|_{g, \pi=\text{constant}}
\ee
Applying Eq. \ref{weinbergeq}, we find $L|_{g, \pi=\text{constant}}=0$, which fine tunes the potential $V(\pi_1, \pi_2)$ to be vanishing in the Minkowski vacuum, ruling out a solution to the cosmological constant problem by dynamical adjustment of the fields.

One might hope to promote the self tuning theories in the bi-galileon model to fully covariant theories that evade Weinberg's no go theorem.  Although it is conceivable that self tuning is spoilt by covariantisation, one ought to recover it in the decoupling  limit $M_{pl} \to 0$. Assuming for the moment that it is not spoilt, it is clear that we evade Weinberg's no go theorem by breaking Poincar\'e  invariance. On the self tuning background, the physical metric is certainly Minkowski but the scalar fields are not all constant, rather $\pi \propto x_\mu x^\mu$. Of the Poincar\' e symmetries, only translational invariance is broken, while Lorentz invariance is preserved.

\subsection{Spherically symmetric solutions and the breakdown of the galileon description} \label{sphback}
Of course, we do not live in a vacuum, and so it is important to ask what happens when we introduce a heavy source into our system. Here we are interested in spherically symmetric excitations  of the self tuning vacua around the Sun. As in the self accelerating scenario, we need some mechanism to suppress modifications of GR at short scales. Here this corresponds to the break down of linearised theory through the Vainshtein mechanism. The equations that govern the excitations are the same as those given in section \ref{sec:selfacc:vain}, by Eqs. \ref{ssceom1}, \ref{ssceom2}, \ref{sph1} and \ref{sph2}.

However, it turns out that although one can engineer a Vainshtein effect at the level of these equations around self tuning vacua, one cannot do so without introducing a large amount of backreaction and destroying the galileon description altogether. To see this most efficiently, we will  present an heuristic argument that illustrates the problem succinctly. Recall that the scalars will backreact heavily on to the geometry unless 
\be \label{ineq}
\left|T_\textrm{scalar}^{\mu \nu}[\eta; \pi, \xi]\right| \ll M_{pl}^2 \left| {\cal E} h^{\mu\nu} \right|
\ee where $h^{\mu\nu}=\tilde h^{\mu\nu}+2\pi \eta^{\mu\nu}$ is the {\it physical} metric perturbation and  
$$
T_\text{scalar}^{\mu\nu}[\eta; \pi, \xi]=\left[\frac{2}{\sqrt{-g}} \frac{\delta }{\delta g_{\mu\nu}}\int d^4 x \sqrt{-g}~{\cal L}_\text{scalar}
\right]_{g_{\mu\nu}=\eta_{\mu\nu}}
$$ 
Here ${\cal L}_\text{scalar}[g; \pi, \xi]$ is constructed out of the covariant completion of $\hat{ \cal  L}_{\pi, \xi}={ \cal  L}_{\pi, \xi}-3M_{pl}^2 \pi \pd^2 \pi$ , as described in \cite{otherpaper}. The full set of galileon equations can be expressed as
\be \label{galeqs}
\textrm{E}_\pi [\pi, \xi]=\frac{\delta}{\delta \pi} \int d^4 x ~ {\cal L}_{\pi, \xi}=-\eta_{\mu\nu} T^{\mu\nu}, \qquad \textrm{E}_\xi[\pi, \xi]=\frac{\delta}{\delta \xi} \int d^4 x ~{\cal L}_{\pi, \xi}=0, \qquad {\cal E} \tilde h^{\mu\nu}=\frac{T^{\mu\nu}}{ M_{pl}^2}
\ee
where $\pi=\bar \pi+\pi_s(r), ~\xi=\bar \xi+\xi_s(r)$ correspond to the scalars evaluated on the spherically symmetric excitation about the self tuning vacuum, and $\tilde h^{\mu\nu}=\tilde h^{\mu\nu}_\textrm{vac}+\tilde h^{\mu\nu}_s$ is the corresponding graviton. The energy-momentum tensor has two pieces: a large vacuum energy and the contribution from the Sun,
\be
T^{\mu\nu}=-\sigma \eta^{\mu\nu}+\delta T^{\mu\nu}_\odot
\ee
For self tuning vacua, the physical metric only contains a contribution from the spherically symmetric excitation $h^{\mu\nu}=h^{\mu\nu}_s=\tilde h^{\mu\nu}_s+2\pi_s \eta^{\mu\nu}$. This is because the vacuum contribution to the physical metric vanishes on account of $\tilde h^{\mu\nu}_\textrm{vac}$ being cancelled by $2\bar \pi \eta^{\mu\nu}$. Furthermore, in the event of a successful Vainshtein mechanism, the graviton excitation should dominate the scalar at short distances, so we have $h^{\mu\nu} \approx \tilde h^{\mu\nu}_s$.  Since the equation for the graviton excitation is really just ${\cal E}\tilde h^{\mu\nu}_s=\frac{\delta T^{\mu\nu}_\odot}{M_{pl}^2}$,  it follows that at short distances, below the Vainshtein scale,
\be
M_{pl}^2 \left|{\cal E}h^{\mu\nu}\right|=\left|\delta T^{\mu\nu}_\odot \right|
\ee
This fixes one half of the inequality \ref{ineq},  governing backreaction. We now turn our attention to the other half, by first noting that,  as a result of diffeomorphism invariance,
\be \label{divT}
\pd_\mu T_\textrm{scalar}^{\mu \nu}[\eta; \pi, \xi]=  \hat{ \textrm{E}}_\pi \pd^\nu \pi+ \hat{ \textrm{E}}_\xi \pd^\nu \xi
\ee
 where 
 $$\hat{\textrm{E}}_\pi =\frac{\delta}{\delta \pi} \int d^4 x ~\hat {\cal L}_{\pi, \xi}=\textrm{E}_\pi-6M_{pl}^2 \Box \pi, \qquad \hat{\textrm{E}}_\xi =\frac{\delta}{\delta \xi} \int d^4 x ~\hat {\cal L}_{\pi, \xi}=\textrm{E}_\xi
 $$
 From the galileon equations of motion (\ref{galeqs}), it follows that {\it  on shell}:  $\hat{\textrm{E}}_\pi=-\eta_{\mu\nu} \delta T^{\mu\nu}_\odot-6M_{pl}^2 \Box \pi_s$ and $\hat{\textrm{ E}}_\xi=0$. In the Vainshtein region, linear contributions are subleading, and so one can neglect the linear $\pi_s$ term in $\hat{\textrm{E}}_\pi$, and approximate it as $\hat{\textrm{E}}_\pi\approx -\eta_{\mu\nu} \delta T^{\mu\nu}_\odot$. Now plugging all of this into Eq. \ref{divT}, we find that 
 $$
 \pd_\mu T_\textrm{scalar}^{\mu \nu}[\eta; \pi, \xi] \approx -\eta_{\alpha\beta } \delta T^{\alpha \beta}_\odot \pd^\nu \bar \pi
$$
Note that we have used the fact that $\bar \pi \gg \pi_s$. This is certainly true except at extremely short distances of the order $|x| \lesssim M_{pl}/\sqrt{|\sigma|}$. For a large vacuum energy, this short distance cut-off is tiny, corresponding to around a millimetre for a TeV scale vacuum energy. 

In terms of scale, our analysis suggests that
\be
\left| T_\textrm{scalar}^{\mu \nu}[\eta; \pi, \xi] \right| \sim |\bar \pi| \left|  \delta T^{\mu\nu}_\odot \right|
\ee
For the inequality (\ref{ineq}) to hold, it is clear that we must have $|\bar \pi| \ll 1$. However, recall that $\bar \pi=-\frac{\sigma}{12 M_{pl}^2} x_\mu x^\mu$, which means that $\bar \pi$ is large on solar system scales for large vacuum energy.  For TeV scale vacuum energy, this suggests we have a big problem with backreaction at solar system scales. In contrast, for $\sigma\sim (meV)^4$, the backreaction is small on subhorizon $|x|<H_0^{-1}$, although there is little phenomenological motivation to self-tune such a small vacuum energy.

Our conclusion is that one can self tune away the vacuum energy in the bi-galileon model; but if the vacuum energy is very large (e.g.~TeV scale or larger) as predicted by current particle theories, it is impossible to do so without abandoning either the Vainshtein effect, or the galileon description, in the vicinity of the solar system. Of course, our arguments are suggestive rather than precise, so one might wish for a subtle resolution of this problem. We are, however, pessimistic. We have been unable to find a numerical example that does {\it not} behave in precisely the way suggested by our heuristic argument.

\section{Discussion} \label{sec:discuss}
We have studied interesting phenomenological solutions to the bi-galileon model \cite{otherpaper}. This can be understood as the decoupling limit of a gravity theory in which GR is modified  by the addition of two scalar fields taking part in the gravitational interaction.  This model is an extension of the original galileon model \cite{gal} to two galileon fields, and is expected to be particularly relevant to co-dimension 2 braneworlds \cite{otherpaper}.

We have focused on two particular types of solution: asymptotically self-accelerating solutions and asymptotically self tuning solutions. Let us first comment on self acceleration. In contrast to the single galileon case we have shown that one can find bi-galileon theories that do not contain any tadpoles, and admit self accelerating solutions that satisfy each of the following:
\begin{itemize}
\item fluctuations about the vacuum do not contain a ghost
\item spherically symmetric galileon fields undergo Vainshtein screening in the solar system
\item fluctuations about the spherically symmetric galileons are never superluminal
\item fluctuations about the spherically symmetric galileons never lead to trouble with excessive emission Cerenkov radiation.
\item do not have an unacceptably low momentum scale for strong coupling, leading to the breakdown of the classical solution in the solar system due to large quantum fluctuations. 
\item do not suffer from problems with backreaction, leading to the breakdown of the galileon description for either the vacuum solution or the spherically symmetric solution.
\end{itemize}
It is, perhaps, remarkable that we can {\it simultaneously}  achieve each of the above in a given model, in contrast to what could be achieved for the case of a single galileon. We believe this merits much more investigation into the bi-galileon model as an alternative to dark energy. 

Of course, the first step is to promote a good theory to a fully covariant one. There are two ways in which we might think about doing this. The first is to simply take our effective $4D$ theory and perform a covariant completion, along the lines described in \cite{covgal,pforms}.  Although, the Galilean invariance is broken, we expect the generic features of the galileon solution to be retained, at least up to corrections which are Planck suppressed. An alternative, more ambitious, approach would be to try to  oxidise our theory and interpret it as a particular co-dimension  two braneworld model with very desirable properties. To this end we note that the probe DBI brane description \cite{dbigal} was very recently generalised to the case of multi-galileons \cite{wesley}.

One might reasonably ask how natural our ``good" theories are? How stable are they against radiative corrections? Radiative corrections  will typically come from two different sources: (i)  galileon loops that  will renormalise the coefficients in the action; and (ii) matter loops that  can potentially introduce non-galilean invariant terms as the coupling to matter breaks the Galilean symmetry. If we treat the galileon theory as an effective theory valid up to the strong coupling scale, $\Lambda_{sc}$, we do not see any problem with naturalness arising from galileon loops. In contrast, matter loops are potentially more dangerous, as the effective theory for the matter sector is valid up to nearly a TeV.  Of course, this is at the origin of the old cosmological constant problem, and its resolution is beyond the scope of this paper.

This brings us nicely on to the other kind of solution studied in this paper: the self-tuning solution. By breaking Poincar\'e invariance one can escape the clutches of Weinberg's no theorem \cite{nogo}, such that in the presence of a vacuum energy, the scalars adjust themselves accordingly, and eliminate the resulting curvature.  The problems start when one requires self-tuning of a large vacuum as predicted by current particle theories and tries to study spherically symmetric solutions sourced by the Sun. Although it is possible to engineer Vainshtein screening at the level of the galileon description, one cannot do so without the galileon description itself breaking down due a large amount of backreaction of the scalars onto the geometry. We anticipate that this will make it extremely difficult to satisfy solar system constraints in a covariant completion of our self tuning galileon models. 

Intuitively, this is actually quite easy to understand. On cosmological scales, we are asking the galileon fields to do an awful lot of work in screening the large vacuum energy from the resulting curvature. Indeed, it requires a background galileon field, $\bar \pi|_\textrm{self-tun} \gg 1$ at Hubble distances, in contrast to self-acceleration which has  $\bar \pi|_\textrm{self-acc} \sim 1$ at Hubble distances. At the same time we are asking that the galileon fields {\it do nothing}  on solar system scales, that they are screened by the Vainshtein mechanism and one is able to recover GR. It seems that such a dramatic change in the galileon behaviour between cosmological and solar system scales is impossible to achieve.

As we stated in the introduction, the Vainshtein mechanism is not the only means of suppressing scalar fields in the solar system. If we are prepared to break Galilean invariance in the vacuum theory then we might consider an alternative mechanism whereby the scalars develop a large mass in the vicinity of heavy objects like the earth. Such mechanisms include the chameleon \cite{cham} and the symmetron \cite{symm}, and we could even consider using them in tandem with the Vainshtein mechanism.  Even so,  as described in the previous paragraph,  we are asking the scalar fields to change their behaviour dramatically, perhaps {\it too} dramatically.  By making use of chameleons, we might also have to worry about possible violations of the Equivalence Principle \cite{ep}. In the event of an unsuccessful resolution of these issues, it would be  worth asking if Weinberg's no go theorem \cite{nogo} can be extended to allow the breaking of Poincar\'e invariance, but taking into account phenomenological constraints coming from local gravity tests \cite{gravtest}. 

\acknowledgements
We would like to thank Ed Copeland, Stanley Deser,  Alberto Nicolis and Yi Wang for very useful discussions. AP is funded by a Royal Society University Research Fellowship and SYZ by a CSRS studentship.

\section{Erratum}
Some time  after the original publication of our work, it has been argued that it is in fact not possible to simultaneously meet all the consistency criteria required for a pathology-free bigalileon theory with self-acceleration  \cite{c>1}.  We do not attempt to summarize the details of that work, but give a generic parallel argument that agrees with the spirit of their conclusions. If a pathlogy-free theory existed, our goal would be to find a  choice of $a_{m,n}' $ and $b_{m,n}'$ that meets all the relevant consistency criteria spelt out in the main body of the paper. To this end we immediately impose $a_{m, 2-m}'=b_{m, 2-m}'=0$, and  $a_{m, 4-m}'=b_{m, 4-m}'=0$. This ensures that $\Sigma_2=0$ and $\Sigma_4=0$, the former motivated by the need to avoid superluminality at large distances ($\left.\Sigma^{(l)}_2\right|_{r \gg r_V}=\left.\Sigma^{(nl)}_2\right|_{r \gg r_V}=0$), the latter motivated in order to avoid issues with strong coupling at shorter distances, as outlined in the draft. In what follows it will be instructive to note that our spherically symmetric field equations (\ref{ssceom3}), (\ref{ssceom4}) can be succintly written as
\be \label{sigstuff}
\left(\Sigma_1 +\Sigma_2 +\frac{2}{3} \Sigma_3 \right) \left(\begin{array}{c }y\\ z \end{array}\right)= \left(\begin{array}{c } M/4\pi r^3 \\ 0 \end{array}\right)
\ee
We begin by studying the short distance physics in more detail. The first thing to note is that the leading order field equations require that $\frac{2}{3} \left.\Sigma_3^{(l)}\right|_{r \ll r_V} \left(\begin{array}{c }y^{(l)}\\ z^{(l)} \end{array}\right)= \left(\begin{array}{c } M/4\pi r^3 \\ 0 \end{array}\right)$, yielding $y^{(l)}\sim z^{(l)} \sim 1/r$ at short distances. However, we previously  argued that to avoid problems with Cerenkov radiation due to the earth's motion through a slowly propagating $\pi$ field, one ought to take $\left. \Sigma_3^{(l)}\right|_{r \ll r_V}=0$. Having rewritten the field equations in the suggestive form given by equation (\ref{sigstuff}), we now see that such a condition would alter the scaling of $y^{(l)}$ and $z^{(l)}$ with radius, and one would need to reanalyse the behaviour. However, one may take the view that we need not worry too much about a slowly propagating scalar well within the Vainshtein radius. This is because the Vainshtein screening renders the scalar weakly coupled to matter, so this ought to suppress the rate of Cerenkov emission. Thus we take a relaxed view of this, and allow $\left. \Sigma_3^{(l)}\right|_{r \ll r_V} \neq 0$, while asking if all other remaining pathologies may be avoided.

To this end, we stay at short distances. Since $\left. \Sigma_3^{(l)}\right|_{r \ll r_V} \neq 0$, we have ${\cal M} \approx \cos^2 q I$, where the angle $q$ is defined just after equation (\ref{disp}). This gives $c^2(q)=\cos^2 q$, ensuring no superluminality for sufficiently large $q$.  For very small $q$ one must study the next to leading order behaviour to see if we are pushed towards $c^2>1$. In \cite{c>1} it is shown in detail how such a scenario can and does occur if one also  requires the absence of ghost. We will not reproduce that argument here,  showing instead how pathologies  arise elsewhere as well.

The absence of ghost at short distances now requires $\Sigma_3^{(l)}$ to have positive eigenvalues. With very little loss of generality, let us assume that $y^{(l)}=z^{(l)}\sim 1/r$ in this regime. The second part of (\ref{sigstuff}) now implies that 
\be
b_{0, 3}'=-3 a_{0, 3}'-a_{1,2}'-\frac{1}{3} a_{2,1}'
\ee
 and one can further show that the ghost is absent  at short distances provided
\bea
&&X=a_{0, 3}'+a_{1, 2}'+a_{2, 1}'+a_{3, 0}'>0 \\
&&Y=3a_{0, 3}'+2a_{1, 2}'+a_{2, 1}'<0
\eea

We now turn to the large distance behaviour, where we now assume that $y^{(l)}=Rz^{(l)} \sim 1/r^3$, for some real constant $R$. To avoid a ghost in this regime we require $\Sigma_1$ to have positive eigenvalues. Since $\Sigma_1 \left(\begin{array}{c }y^{(l)}\\ z^{(l)} \end{array}\right)= \left(\begin{array}{c } M/4\pi r^3 \\ 0 \end{array}\right)$ in this regime, we deduce that $b'_{0,1}=-Rb_{1, 0}'$, and note that the ghost may be avoided as long as  we take $a'_{1,0}>| b'_{1,0}|/|R|$ and $sgn(b'_{1,0})=-sgn(R)$. This is obviously easily achieved. 

It remains to ask if we may also avoid superluminality at large distances. This requires that $\left. \Sigma_3^{(l)}\right|_{r \gg r_V}$ has negative eigenvalues. Defining 
\bea
&& Q_1(R)=a_{1, 2}'+2a_{2, 1}'R+3a_{3, 0}'R^2 \\
&& Q_2(R)= 3a_{0, 3}'+2a_{1, 2}'R+a_{2, 1}'R^2 \\
&& Q_3(R)=  3b_{0, 3}'+6a_{0, 3}'R+a_{1, 2}'R^2
\eea
this condition is equivalent to demanding that $Q_1<0, Q_3<0$ and $Q_1Q_3>Q_2^2$. However,  let us assume that all but the last criteria are met, setting $X=s^2, Y=-3t^2, Q_1(R)=-u^2, Q_3^(R)=-v^2$,  where $s, t, u$ and $v$ are real. We then find that $Q_2(R)=\frac{3}{2}(s^2+t^2)R^2-3Rt^2+\frac{3}{2}t^2+\frac{1}{2}(u^2+v^2)>\frac{1}{2}(u^2+v^2)+\frac{3}{2}\frac{s^2t^2}{s^2+t^2}>| u v |=\sqrt{Q_1(R) Q_3(R)}$. Thus we are unable to satisfy the condition that $Q_1Q_3>Q_2^2$, suggesting the presence of superluminal modes. Of course, this long distance superluminality can be avoided if we relax our no ghost condition at short distances by allowing the parameters $s$ or $t$ to take on imaginary values. Perhaps this  scenario is more accurately identified as a strong coupling issue rather than one of a ghost. This is because there is no ghost at large distances in these cases, so in order for the fluctuations to become ghostlike at short distances, the coefficient of the kinetic term must have passed through zero determinant. This corresponds to a region of strong coupling beyond which we can no longer trust our classical solution.

\end{document}